\documentclass{moriond}

\def\be{\begin{equation}}
\def\ee{\end{equation}}
\def\bea{\begin{eqnarray}}
\def\eea{\end{eqnarray}}

\newcommand{\Photo}{}

\begin{document}
\vspace*{4cm}
\title{TOWARDS HOLOGRAPHIC QCD:\\ AdS/CFT, CONFINEMENT DEFORMATION, AND DIS at SMALL-X}

\author{ R. C. BROWER }

\address{Boston University, Boston MA 02215, USA}

\author{M. DJURI\'C}

\address{Universidade do Porto, 4169-007 Porto, Portugal}

\author{T. RABEN and C.-I TAN}

\address{Brown University, Providence, RI 02912, USA}

\maketitle\abstracts{
We investigate the softwall AdS/CFT model.~\cite{Brower2014}  We specifically looked at the Pomeron, leading Regge contribution to a scattering process and used it to fit deep inelastic scattering data from the HERA collaboration.  We find that the model fits the data with much more success than the purely conformal case, and find similar success to previous confinement models.}

\section{Introduction}
AdS/CFT has passed many serious tests and does an excellent job of describing a four-dimensional strongly coupled conformal field theory.  Shortly after its inception, it was proposed that this correspondence could be used to describe QCD physics.  Specficially, by deforming the string theory, the conformal gauge theory would develop confinement behavior. ~\cite{Maldacena1998,Polyakov1998}  One of the early successes of the AdS/CFT was that the geometric scaling of the AdS theory could soften the historically troublesome energy dependence of high energy string scattering.~\cite{Polchinski2002}  This in turn indicated that the gauge gravity correspondence might be able to be used for physical processes, like deep inelastic scattering (DIS), where strongly coupled physics plays an important role.~\cite{Polchinski2003}  In this holographic picture, glueballs could be described~\cite{Brower2000b} and the AdS Pomeron was unambiguously identified as the Regge trajectory of the graviton.~\cite{Brower2007} The strongly coupled dynamics of this Pomeron and its eikonalization were identified~\cite{Brower2009a}, and then these techniques were extended to the AdS Odderon.~\cite{Brower2009} Pomeron exchange in AdS was then applied to fit small-x HERA data for DIS, DVCS and vector meson production.~\cite{Brower2010,Costa:2012fw,Costa:2013uia}

In this paper, we consider the graviton fluctuations of type IIB string theory in a compactified AdS$_5\times$S$^5$ background, which is geometrically deformed with a soft (gradual) confinement.
\be
ds^2= \frac{R^2}{z^2}\left[dz^2+dx\cdot dx\right]+R^2d\Omega_5 \rightarrow e^{2A(z)}\left[dz^2+dx\cdot dx\right]+R^2d\Omega_5
\ee

\noindent Here R is the radius of both spaces, x is a usual 4-dimensional minkowski coordinate, and z is the AdS radial direction.  For a \emph{purely geometric} softwall confinement, the scaling function can be identified as $A(z)=\Lambda^2z^2+$ln$(R/z)$.  Where $\Lambda$ will set the confinement scale.

We examined a DIS process, where a lepton scatters from a proton via the exchange of a virtual photon. In terms of the virtual Compton subprocess, we specifically examined the Regge limit: s$\approx$Q$^2$/x large and Q$^2$ fixed.  In this so-called small-x limit, confinement effects will play a particularly imprtant role.  We will obtain physical results via the optical theorem, where the forward limit of the virtual photon scattering will tell us about the total cross section, $\sigma_{total}=\frac{1}{s}Im\left[\mathcal{A}(s,t=0)\right]\sim \frac{1}{s}Im\left[\chi(s,t=0)\right]$ The total cross section can then be used to fit one of the hadronic structure functions, F$_2$\footnote{The hadronic tensor, the hadronic contribution to a scattering amplitude, can be written in terms of several structure functions.  In certain kinematic regimes, the structure functions can be written in terms of each other, and thus there a determination of the F$_2$ behavior determines the hadronic behavior of the scattering process.}, from the combined HERA and ZEUS experiments.  Explicitly,\hspace{5pt}$
F_2(x,Q^2) = \frac{Q^2}{4\pi^2\alpha_{em}}\left(\sigma_{trans}+\sigma_{long}\right)$

\section{SoftWall Model}
The soft wall model was originally proposed in~\cite{Karch2006a}.  It showed what type of AdS confinement would lead to linear meson trajectories.  Several dynamical softwall toy models, where the confinement is due to a non-trivial dilaton field, have subsequently been described.~\cite{Batell2008b} There has even been some success in using the softwall model to fit QCD mesons.~\cite{Katz2006}  \footnote{These models involve a dynamical dilaton and tachyon field, but there is still debate about some of the signs of some parameters.~\cite{Karch2011,Teramond2010c}  However, for our pourpuse, we only need to consider graviton fluctions to describe the AdS Pomeron.  For the dynamical soft wall models, the graviton does \emph{not} couple to the dilaton field--and thus a purely geometric confinement model is sufficient to consider.} Significant effort has been put forth to develop standard model and QCD features in these softwall models.~\cite{Batell2008c,Csaki2007,Erlich2005}

%The softwall propagator can be shown to obey ~\cite{Brower2014}

%\be
%\left[-\partial_z^2+10\Lambda^2+4\Lambda^4z^2-t+\frac{12-\alpha^2(j)R^5}{z^2}\right]\chi_P(j,z,z',t)=\delta(z-z')
%\ee

In the softwall model, the graviton dynamics involve a spin dependent mass-like term $\alpha^2(j)=2\sqrt{\lambda}(j-j_0)$.  The Pomeron propagator can take several forms: for quantized momentum transfer, $t\rightarrow t_n$, the solution behaves like Laguerre polynomials: $\chi\sim L^{\alpha}_n(2\Lambda^2z^2)$.  More generally, for a continuous t spectrum, the solution is a combination of Whittaker's functions

\be
	\chi_P(j,z,z',t)=\frac{M_{\kappa,\mu}(z_<)W_{\kappa,\mu}(z_>)}{W(M_{\kappa,\mu},W_{\kappa,\mu})}
\ee

for $\kappa=\kappa(t)$ and $\mu=\mu(j)$ .  $\Lambda$ controls the strength of the soft wall and in the limit $\Lambda \rightarrow 0$ one recovers the conformal solution\footnote{This has a similar behavior to the weak coupling BFKL solution where Im$(\chi(p_{\perp},p_{\perp}',s))\sim\frac{s^{j_0}}{\sqrt{\pi \mathcal{D}ln(s)}}exp(-(ln (p_{\perp}')-ln(p_{\perp}))^2/\mathcal{D}ln(s))$}\hspace{5pt}$	Im(\chi_P^{conformal}(t=0))=\frac{g_0^2}{16}\sqrt{\frac{\rho^3}{\pi}}(zz')\frac{e^{(1-\rho)\tau}}{\tau^{1/2}}exp\left(\frac{-(ln(z)-ln(z'))^2}{\rho\tau}\right)$

If we look at the energy dependence of the pomeron propagator, we can see a softened behavior in the regge limit.  In the forward limit, $t=0$, the conformal amplitude scales as $-s^{\alpha_0}log^{-1/2}(s)$, but this behavior is softened to $-s^{\alpha_0}log^{-3/2}(s)$ in the hardwall and softwall models  This corresponds to the softening of of a j-plane singularity from $1/\sqrt{j-j_0}\rightarrow\sqrt{j-j_0}$.

\section{Numerics}
The data examined comes from the combined H1 and ZEUS experiments at HERA.~\cite{Aaron2010a} A fit was done with the same methods used previously for the conformal and hardwall models in~\cite{Brower2010}, making the results directly comparable. 

\begin{figure}[ht]
\begin{minipage}{0.5\linewidth}
\centerline{\includegraphics[width=0.7\linewidth]{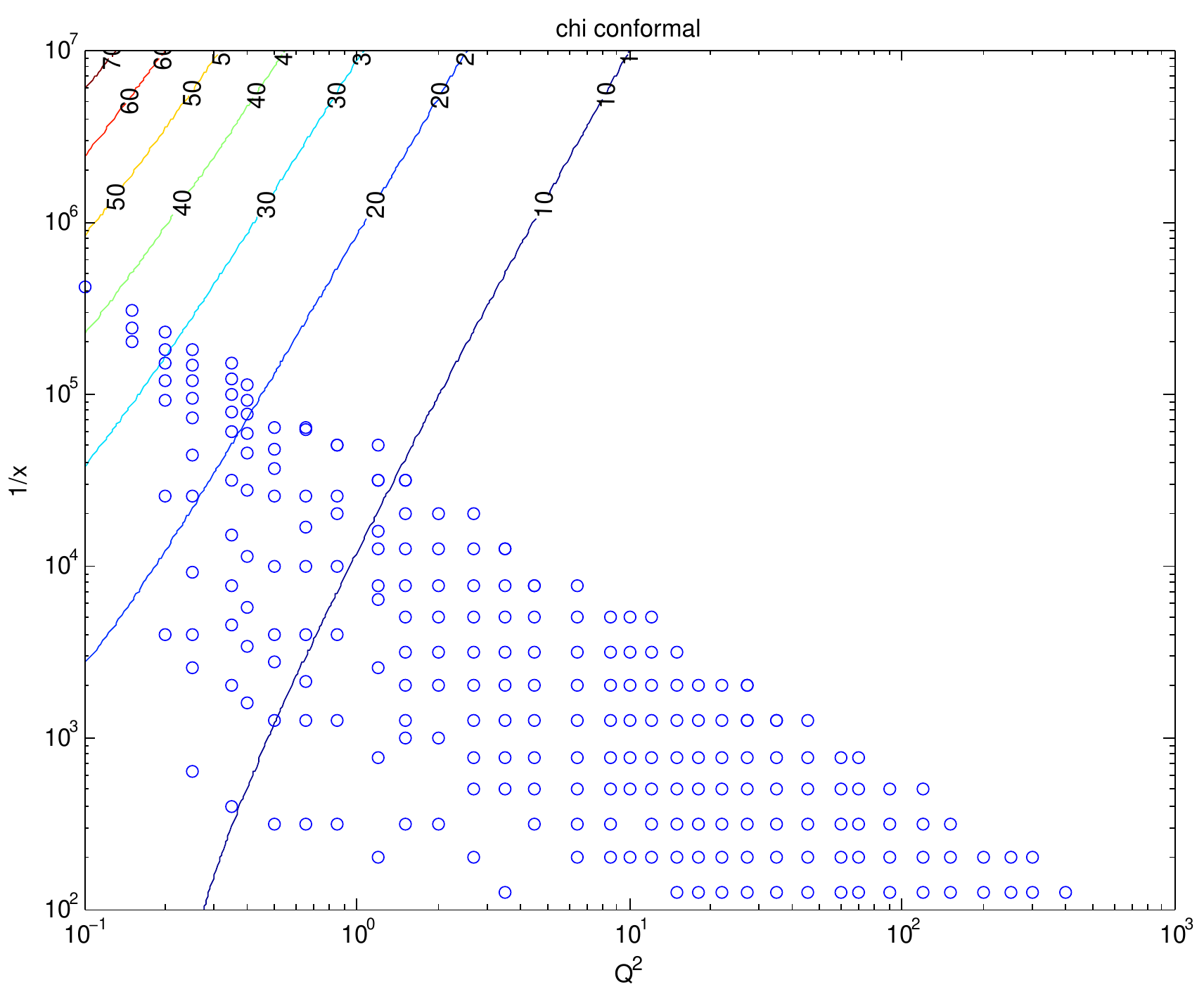}}
\end{minipage}
\hfill
\begin{minipage}{0.5\linewidth}
\centerline{\includegraphics[width=0.7\linewidth]{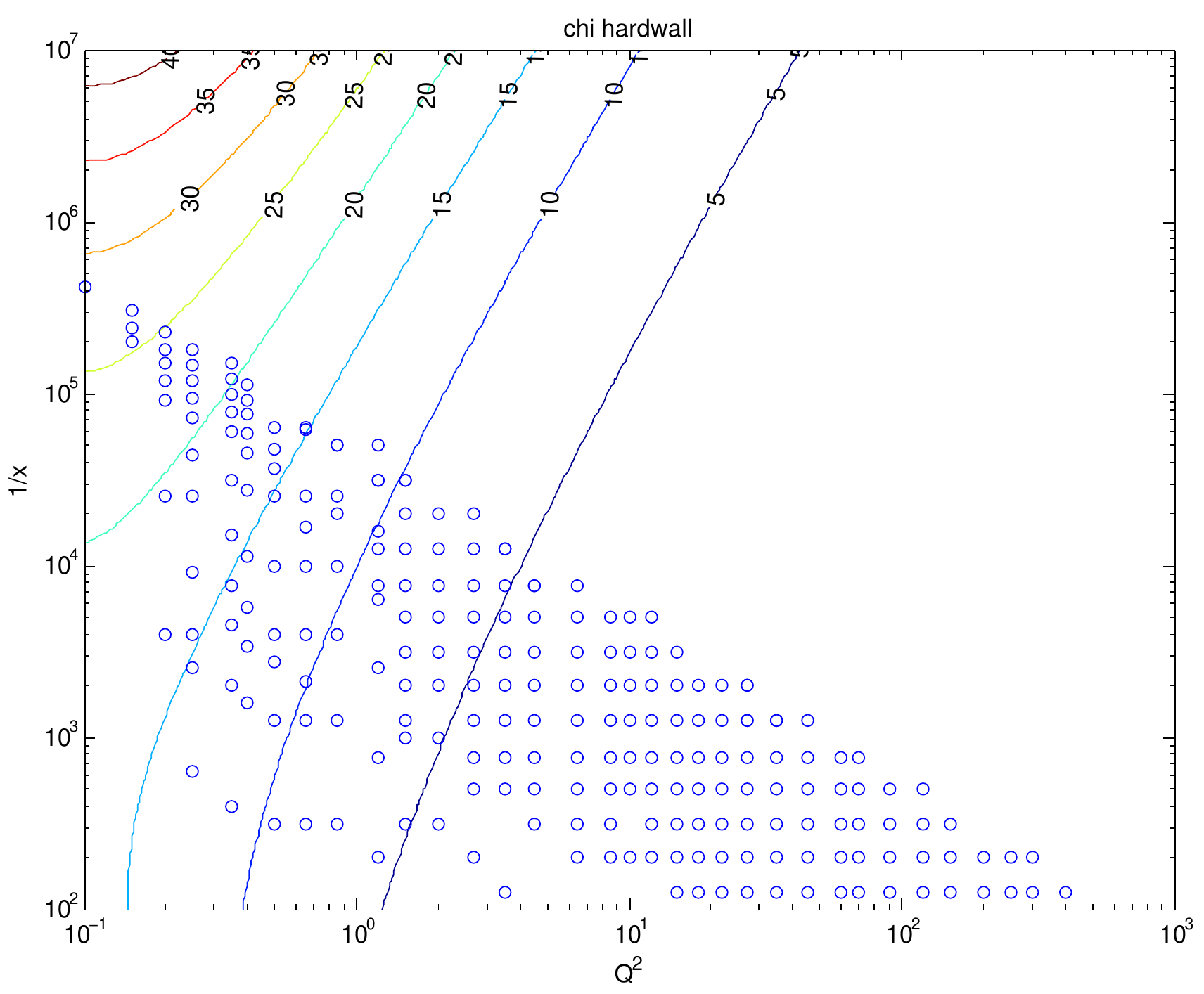}}
\end{minipage}
\end{figure}

\begin{figure}[ht]
\begin{minipage}{0.5\linewidth}
\centerline{\includegraphics[width=0.7\linewidth]{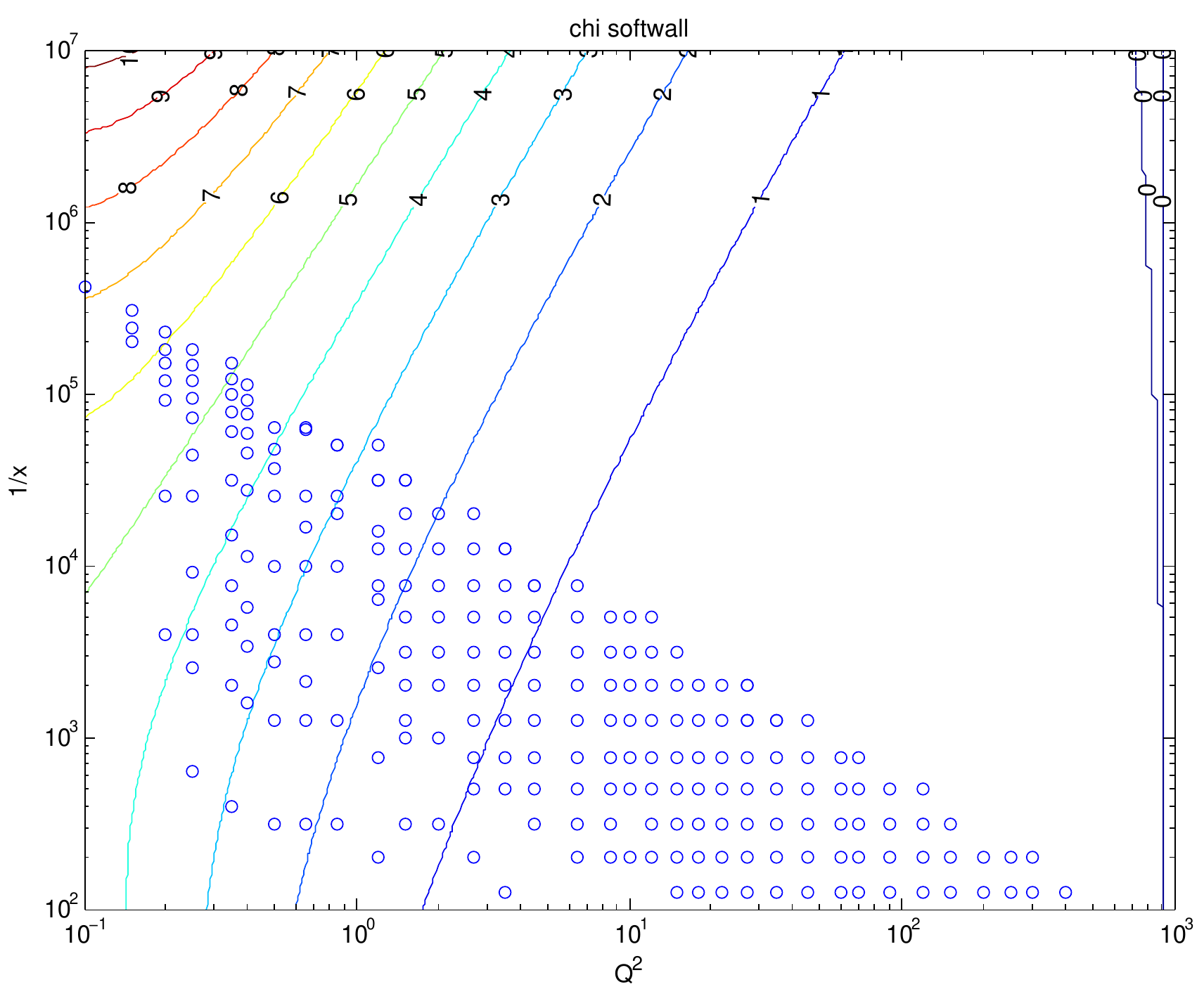}}
\end{minipage}
\hfill
\begin{minipage}{0.5\linewidth}
\centerline{\includegraphics[width=0.7\linewidth]{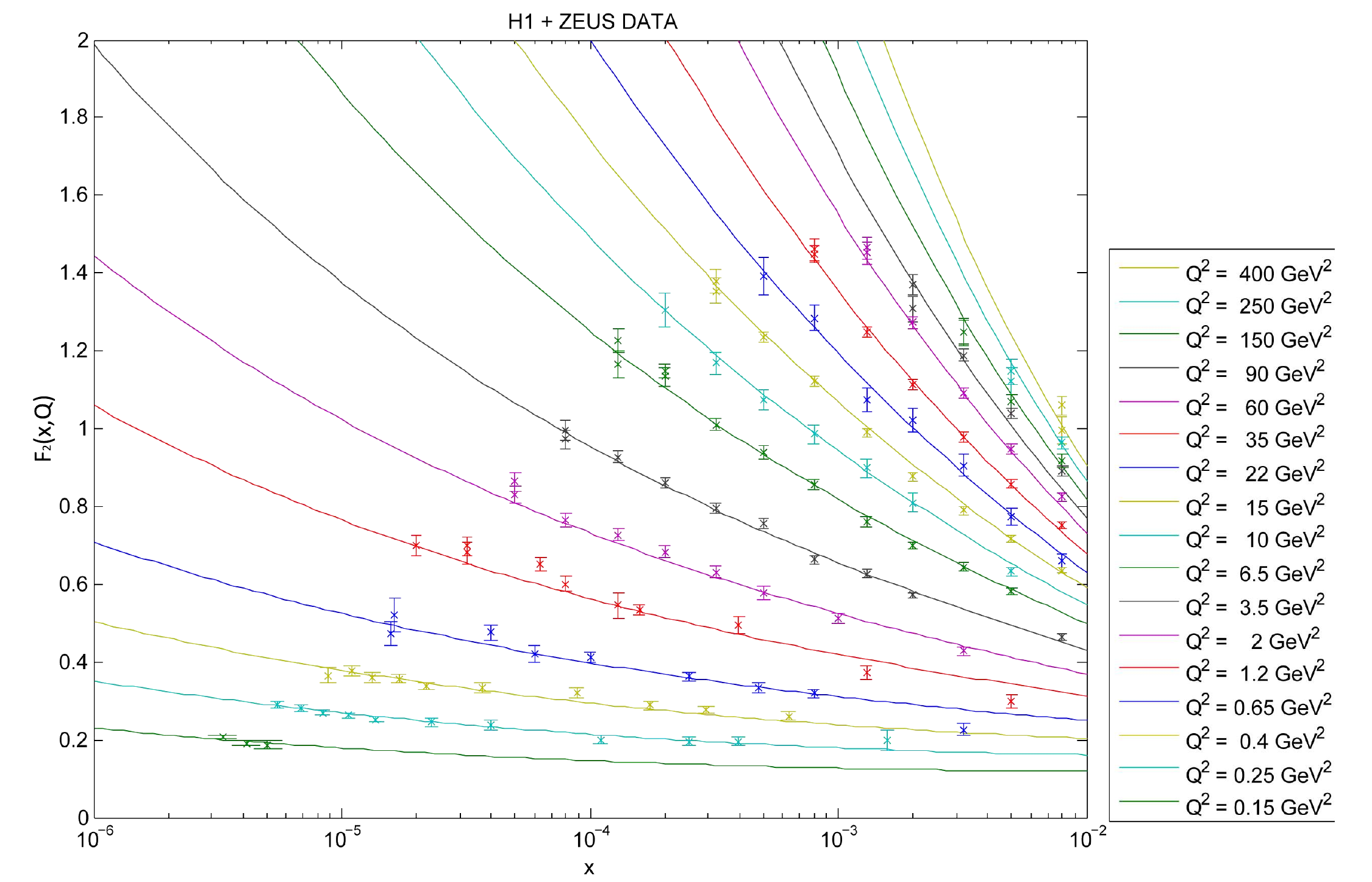}}
\end{minipage}
\hfill
\caption[]{Contour plots of Im($\chi$) for the conformal (top left), hardwall (top right), and softwall (bottom left) models.  The softwall was also used to fit the F$_2$ proton structure function (bottom right).}
\label{fig:chi}
\end{figure}

\begin{table}[hb]\footnotesize
\centering
	\begin{tabular}{|c||c|c|c|c|c|}
	\hline
	Model & $\rho$ & $g_0^2$ & $z_0$ (GeV$^{-1}$) & Q' (GeV)& $\chi^2_{dof}$ \\ \hline
	conformal & $0.774^*\pm$0.0103 & $110.13^*\pm1.93$ & -- & $0.5575^*\pm0.0432$ & 11.7 $(0.75^*)$ \\ \hline
	hard wall & $0.7792\pm0.0034$ & $103.14\pm1.68$ & $4.96\pm0.14$  & $0.4333\pm0.0243$  & 1.07 $(0.69^*)$ \\ \hline
	softwall & 0.7774 & 108.3616 & 8.1798  & 0.4014  & 1.1035 \\ \hline
	softwall*& 0.6741 & 154.6671 & 8.3271  & 0.4467 & 1.1245\\ \hline
	\end{tabular}
	\caption[]{	Comparison of the best fit (including a $\chi$ sieve) values for the conformal, hard wall, and soft wall AdS models.  The final row includes the soft wall with improved intercept.}
	\label{tab:fit}
\end{table}

The softwall* row describes indicates that the fit was run using a pomeron intercept (which determines $\lambda$) up to order $\mathcal{O}(\lambda^{-5/2})$.~\cite{Brower2013}  This quantity has been calculated to high order using integrability  and Regge techniques in $\mathcal{N}=4$ SYM~\cite{Basso2011,Gromov2014b,Kotikov2013,Costa2012a}.  

\section{Conclusions}
The softwall model continues to fit the known DIS data extremely well.  The fits all had similar success to that of the previously investigated hardwall model.  In both cases, the models lead to an \emph{extremely} better fit than the conformal case, indicating that at the considered x and Q scales, confinement plays an important role.

There are still things left to investigate for the softwall model.  The propagator in general can be solved to higher order in j.  This would in principle improve the accuracy, but it requires doing a difficult string calculation.  Also, the details of describing mesons and other composite particles is still not completed.  Immediately however, the current softwall model can still be applied to various situations.  In the limit $t\rightarrow10\Lambda^2$ the equations of motion greatly simplify and the model reduces to a $1+1$ dimensional conformal model where CFT techniques might be able to improve understanding.~\cite{Alfaro1976}

\section*{Acknowledgements}
The works of C-I.T. and T.R. were funded by DE-SC0010010-Task A. The work of M.D. was partially funded by grants PTDC/FIS/099293/2008 and CERN/FP/116358/2010 and supported by the FCT/Marie Curie Welcome II program.  The authors would like to thank Miguel Costa for many beneficial talks on this and related works.
\section*{References}

%\begin{thebibliography}{99}
\bibliography{softwallrefs}
\end{document}